\documentstyle[12pt,fullpage,epsf]{article}

\newcommand{\dint}[1]{\int{d^{#1}k\over (2\pi)^{#1}}}
\newcommand{\br}{{\bf r}}

\begin{document}
\font\ninerm = cmr9


\def\footnoterule{\kern-3pt \hrule width \hsize \kern2.5pt}

\pagestyle{empty}
\begin{center}
{\large\bf Perturbative Analysis of Nonabelian Aharonov-Bohm Scattering}%
\footnote{\ninerm This work is supported
in part by funds provided by the U.S. Department of Energy (D.O.E.)
under contracts \#DE-AC02-76ER03069 and \#DE-AC02-89ER40509, as well as in part
by the National Science Foundation under contracts \#INT-910559 and
\#INT-910653.}

\vskip 1cm
Dongsu Bak and Oren Bergman
\vskip 0.5cm
{\it Center for Theoretical Physics\\
Laboratory for  Nuclear Science and Department of Physics\\
Massachusetts Institute of Technology\\
Cambridge, Massachusetts 02139, U.S.A.}

\end{center}

\vspace{1.2cm}
\begin{center}
{\bf ABSTRACT}
\end{center}

{\leftskip=0.6in \rightskip=0.6in

We perform a perturbative analysis of the nonabelian Aharonov-Bohm problem to
one loop in a field theoretic framework, and show the necessity of contact
interactions for renormalizability of perturbation theory. Moreover at
critical values of the contact interaction strength the theory is finite and
preserves classical conformal invariance.

}
\vfill

\hbox to \hsize{CTP \# 2283 \hfil hep-th/9403104 \hfil February 1994}

\newpage

\pagenumbering{arabic}
\pagestyle{plain}
\section{Introduction}
The nonabelian generalization of the Aharonov-Bohm (AB) effect \cite{ab,ruij}
 is essentially
the scattering of particles carrying nonabelian charge by a tube carrying
a nonabelian magnetic flux. The two body case has recently been solved exactly
\cite{ver,preskill,lee}, by choosing a convenient basis in which the problem
reduces to the abelian AB effect.

Recent interest in the abelian AB effect is due to the fact that anyons
(particles which acquire fractional statistics through the AB effect) are
useful for understanding the Fractional Quantum Hall Effect \cite{fq}, and may
play
a role in High $T_c$ Superconductors \cite{sc}. The exact solution to the AB
scattering
problem has been known for over thirty years, yet it had until recently
\cite{ber} resisted a perturbative treatment. Earlier attempts at a
perturbative solution failed by missing the s-wave contribution in first
order, and producing a divergence in second order \cite{per}. The failure
was explained in Ref.~\cite{exp} by showing that the series expansion of the
exact solution is ill defined for a zero diameter flux tube. In Ref.~\cite{ber}
a field theoretic model for the AB effect was presented. It is based on
Hagen's model \cite{hag1}, but also includes contact interactions. It was shown
in that paper that, for a critical value of the contact
interaction strength, perturbation theory is well defined and gives the correct
conformally invariant
scattering amplitude to one loop. It was also shown that the model possesses
a conformal anomaly away from the critical point. Subsequently, Freedman
et. al. showed that conformal invariance is preserved at this critical
point to three loops \cite{fr}.

The nonabelian generalization of this field theoretic model was first studied
at the classical level in Ref.~\cite{RJ1}. Quantization and the
derivation of the two body Schr\"odinger equation for the nonabelian AB problem
was carried out in \cite{BJP}. So far, a perturbative treatment has not been
attempted, but it is obvious that it will suffer the same difficulties as in
the abelian problem.

The aim of our paper is to perform a perturbative analysis of the nonabelian
AB problem in a field theoretic framework. The field theory we use is a slight
generalization of the one studied in Ref.~\cite{RJ1}. We shall show that the
contact interaction is necessary for renormalizability of the theory, and for
a correct treatment of the AB problem. In section II we introduce the field
theory and review the resulting two body Schr\"odinger equation and its
solution.
In section III we compute the two particle scattering amplitude to one loop and
show that in general renormalization is necessary resulting in a conformal
anomaly. We shall also show that this theory possesses critical points at which
the
anomaly vanishes and conformal invariance is regained. At a particular critical
point, namely the completely repulsive critical contact interaction, the
perturbative scattering amplitude agrees with the exact one.
Section IV is devoted to concluding remarks.

\section{Field Theoretical Formulation}
\renewcommand{\theequation}{\Roman{section}.\arabic{equation}}

Nonrelativistic bosonic particles carrying nonabelian charges are described by
the Lagrange density,
\newpage
\begin{eqnarray}
 {\cal L}&=& -\kappa\epsilon^{\alpha\beta\gamma}{\rm tr}(A_\alpha\partial_\beta
A_\gamma+
 {2g\over 3}A_\alpha A_\beta A_\gamma )+i\phi^\dagger D_t \phi -
 {1\over 2m}({\bf D}\phi)^\dagger\cdot {\bf D}\phi
 \nonumber \\
 &\ &-{1\over 4}\phi^\dagger_{n'}
 \phi^\dagger_{m'}C_{n'm'nm}\phi_{n}\phi_{m} \ ,
 \end{eqnarray}
 where $\phi$ is a complex bosonic field transforming in an irreducible
 representation of the gauge group $G$, generated by the matrices $T_a$ with
 $a=1\ldots dim\, G$, and $A_\mu\equiv A_\mu^a T_a$. The matrices satisfy
 the Lie algebra
 \begin{equation}
 [T_a, T_b]=f_{ab}\,^c T_c \ ,
 \end{equation}
 and are normalized by
 \begin{equation}
 tr\left(T_aT_b\right)=-{1\over 2}h_{ab} \ ,
 \end{equation}
 where $h_{ab}$ is a nonsingular group metric. This metric can be used to raise
 and lower group indices. The covariant derivatives are given by
 \begin{eqnarray}
 D_t&=& \partial_t +gA_0 \\
 {\bf D}&=&\nabla -g{\bf A} \ .
 \end{eqnarray}
 The contact interaction term describes a delta function interaction between
the
 particles.
 Since the particles are bosons we can assume
 \begin{equation}
 C_{n'm'nm}=C_{m'n'mn} \ ,
 \end{equation}
 and from the reality of the Lagrange density, the matrix $C$ should be
 Hermitian,
 \begin{equation}
 C^*_{n'm'nm}=C_{nmn'm'} \ .
 \end{equation}
 To make the notation concise we shall drop the matter indices, and regard four
 indexed objects as components of matrices in the basis $|n, m\rangle$
 , for example,
 \begin{eqnarray}
 C_{n'm'nm}&\equiv &\langle n', m'|{\cal C}|n, m\rangle\nonumber \\
 T^a_{n'n}T^b_{m'm}&\equiv &\langle n', m'|T^a\otimes T^b|n, m\rangle\nonumber
\\
 C^2_{n'm'nm}&\equiv &\sum_{l, l'}\langle n', m'|{\cal C}|l, l'\rangle\langle
l, l'
 |{\cal C}|n, m\rangle \ .
 \end{eqnarray}
 The last definition  in (II.8) agrees with the usual matrix multiplication.
 An additional restriction on the form of ${\cal C}$ comes from gauge
invariance of
 the action,
 \begin{equation}
 [T_a\otimes {\bf 1}+{\bf 1}\otimes T_a,{\cal C}]=0 \ .
 \end{equation}
 By Schur's lemma, $T^2\equiv T_aT^a\propto {\bf 1}$, so
 \begin{equation}
 [T^2\otimes {\bf 1}, {\cal C}]=0 \ \ ,\ \ [{\bf 1}\otimes T^2 , {\cal C}]=0 \
{}.
 \end{equation}
 Using the identity
 \begin{equation}
 T^a\otimes T_a={1\over 2}(T\otimes {\bf 1}+{\bf 1}\otimes T)^2
-{1\over 2}T^2\otimes {\bf 1}
 -{1\over 2}{\bf 1}\otimes T^2 \ ,
 \end{equation}
 we get
 \begin{equation}
 [T^a\otimes T_a, {\cal C}]=0 \ .
 \end{equation}

  To get the most general gauge invariant form of ${\cal C}$,
 let us use a basis that simultaneously
 diagonalizes  $T^a\otimes T_a$, all the other Casimir operators $B$
constructed
 from $T_a\otimes {\bf 1}+{\bf 1}\otimes T_a$, and a maximal set of mutually
 commuting
 operators $W$ chosen from the set of $T_a\otimes {\bf 1}+{\bf 1}\otimes T_a$ :
 \begin{eqnarray}
 T^a\otimes T_a|\alpha, \beta, w\rangle&=&\alpha|\alpha,\beta,
 w \rangle \ .
 \end{eqnarray}
 Here  $\beta$ and $w$ represent the eigenvalues of the operators $B$
 and $W$ respectively. Note  that the matrix ${\cal C}$ is also a Casimir
 constructed from
 $T_a\otimes {\bf 1}+{\bf 1}\otimes T_a$ [cf. (II.9), which holds for {\em all}
 $a$]. Since we already use all
 the Casimir operators in the construction of the basis, the Casimir ${\cal C}$
is
 diagonalized in this  basis and its eigenvalues do not depend on $w$\,:
 \begin{eqnarray}
 {\cal C}|\alpha,\beta, w\rangle&=&c(\alpha,\beta)|\alpha,\beta, w\rangle\ .
 \end{eqnarray}
 Hence the most general gauge invariant form of ${\cal C}$ is given by
 \begin{eqnarray}
 {\cal C}&=&\sum_{\alpha\beta w}
 |\alpha,\beta, w\rangle c(\alpha,\beta)\langle\alpha,\beta, w|\ .
 \end{eqnarray}

 Quantization of this theory in the two particle sector in Coulomb gauge yields
 the following Schr\"odinger equation \cite{lee,BJP}
 \begin{equation}
 i\partial_t \psi (\br_1,\br_2;t)=\left\{\!-{1\over 2m}\left(
 \bigg[\nabla_1\! +\! {ig^2\over \kappa}{\bf G}(\br_1\! -\!\br_2)T^a\!\otimes\!
T_a
 \bigg]
 ^2\!+\!\bigg[1\!\leftrightarrow\! 2\bigg]\right)\! +\! {{\cal C}\over 2}
\delta
(\br_1\!-\!\br_2)\right\}\psi(\br_1, \br_2; t) \ ,
 \end{equation}
 where ${\bf G}(\br)={1\over 2\pi}\nabla \times \ln r$, and in the
 $|n,m\rangle$ basis, $T^a\otimes T_a \psi$ and ${\cal C}\psi $ are
respectively
 \begin{eqnarray}
 (T^a\otimes T_a \psi)_{nm} =(T^a)_{nn'} (T_a)_{mm'}\psi_{n'm'}\nonumber \\
 ({\cal C} \psi)_{nm} ={\cal C}_{nmn'm'}\psi_{n'm'}\ .
 \end{eqnarray}

 The components of the wavefunction in the diagonal basis are given by
 \begin{equation}
 \psi_{\alpha\beta w}=\sum_{n, m}\psi_{nm}\langle n, m|\alpha,\beta,
 w\rangle \ .
 \end{equation}
 In this basis the nonabelian problem is reduced to the
  abelian one. The time independent Schr\"odinger equation in the center of
mass
  frame is
 \begin{equation}
 \left[-{1\over m}
 \bigg(\nabla + 2\pi i\nu{\bf G}(\br)\bigg)^2 +{c\over 2}
 \delta (\br)
 -E\right]
 \psi_{\alpha\beta w}(\br)=0 \ ,
 \end{equation}
 where $\br\equiv \br_1 -\br_2$, and $\nu\equiv -{g^2
 \alpha\over 2\pi\kappa}$.
 With the usual boundary condition $\psi_{\alpha\beta w}(0)=0$ the contact term
 drops out, and the solution
 is \cite{ab,ruij,RJ2}
 \begin{equation}
 \psi_{\alpha\beta w}(r,\theta)=e^{i\left(pr\cos\theta
-\nu({\bar\theta}-\pi)\right)}
 -\sin\nu\pi e^{-i([\nu]+1)\theta} \int^{\infty}_{-\infty}
 {dt\over \pi}e^{ipr\cosh t}{e^{-\{\nu\}t}\over e^{-i\theta}-e^{-t}} \ ,
 \end{equation}
 where $[\nu]$ is the greatest integer part of $\nu$, $\{\nu\}=
 \nu-[\nu]$, and ${\bar\theta}(\br)=\theta(\br)-2\pi n$ when $
 2\pi n \leq \theta < 2\pi(n+1)$. [The overall phase is fixed by the condition
that
 in the partial wave expansion, each ingoing partial wave has the same phase as
the plane
 wave. The vanishing boundary condition at the origin makes the
 delta function in (II.19) irrelevant.]
 The expression is manifestly single valued. The function ${\bar\theta(\br)}$
 is discontinuous along the positive $x$-axis, but the wavefunction is
 continuous.

  The first term in (II.20) is not appropriate as  an incoming wave since
 the discontinuity gives a singular contribution to the particle flux along the
 positive $x$-axis.\footnote{\ninerm This discontinuity of ${\bar \theta}(\br)$
was
not noticed in previous treatments of the AB problem, which led to the
incorrect
conclusion that the phase-modulated plane wave was appropriate as an incoming
wave in the scattering solution\cite{ab,hag2}.}
If we assume a plane wave form for the incident wave, and
 use the identity\footnote{\ninerm Contrary to the statement made by
Hagen\cite{hag2},
the asymptotic relation (II.21) holds by virtue of the fact that
 the scattering matrix for a free particle is given by $S={\bf 1}$. }
 \begin{eqnarray}
 e^{ipr\cos\theta}&=&\sum_{-\infty}^{\infty}(-i)^ne^{in(\theta - \pi)}J_n(pr)
 \nonumber \\
  &{\sim}&\left({2\pi\over ipr}\right)
 ^{1/2}e^{ipr}\delta
 (\theta) \ \ \ \ {\rm as}\  r\rightarrow\infty \ ,
 \end{eqnarray}
 we can cast the solution in the large $r$ limit as
 \begin{equation}
 \psi_{\alpha \beta w}(r,\theta)\sim e^{ipr\cos\theta}+{1\over\sqrt r}
 e^{i(pr +\pi/4)}f_{\alpha}(\theta) \ ,
 \end{equation}
 where
 \begin{equation}
 f_{\alpha}(\theta)=-{i\over \sqrt{2\pi p}}\left[\sin \pi\nu \cot
 {\theta\over 2} -i\sin |\pi\nu| - 4\pi\sin^2{\pi\nu\over 2}
\delta(\theta)\right] \ .
 \end{equation}
 The delta function in the forward direction is crucial for unitarity of the
 scattering matrix \cite{ruij}.

 In the original basis, the c.o.m. scattering amplitude is given by
 \begin{eqnarray}
 f_{n_1n_2\rightarrow n_3n_4}(\theta)&=&\langle n_3, n_4|{\cal F}(\theta)|
 n_1, n_2\rangle \nonumber \\
 {\cal F}(\theta)&=&-{i\over \sqrt{2\pi p}}\left[\sin(\pi{\Omega}) \cot
 {\theta\over 2} -i\sin |\pi{\Omega}| - 4\pi\sin^2{\pi\Omega\over 2}
\delta(\theta)\right] \ ,
 \end{eqnarray}
 where
 \begin{eqnarray}
 {\Omega}&\equiv& -{g^2\over2\pi\kappa}T^a\otimes T_a=-{g^2\over2\pi\kappa}
 \sum_{\alpha\beta w}
 |\alpha,\beta, w\rangle\alpha\langle\alpha,\beta, w|\nonumber\\
 |{\Omega}|&\equiv& {g^2\over2\pi|\kappa|}
 \sum_{\alpha\beta w}
 |\alpha,\beta, w\rangle|\alpha|\langle\alpha,\beta, w|\nonumber\\
 {\Omega}^2&=& {g^4\over4\pi^2\kappa^2}T^aT^b\otimes
T_aT_b={g^4\over4\pi^2\kappa^2}
 \sum_{\alpha\beta w}
 |\alpha,\beta, w\rangle\alpha^2\langle\alpha,\beta, w| \ .
 \end{eqnarray}
 The abelian result is regained if one sets $T={i\over \sqrt{2}}$ and
 $g=\sqrt{2}e$.

 Taking into account the exchange symmetry, the total scattering amplitude is
 \begin{equation}
 f_{n_1n_2\rightarrow n_3n_4}^{tot}=\langle n_3, n_4|{\cal F}(\theta)|
 n_1, n_2\rangle + \langle n_4, n_3|{\cal F}(\theta+\pi)|
 n_1, n_2\rangle \ .
 \end{equation}
 In contrast to the claim in Ref.~\cite{preskill}, the amplitude is
single--valued.
 Let us compare with
 Ref.~\cite{lee} where the scattering amplitude is obtained for $SU(2)$. First
of all,
  our formula
 has a contribution of the delta function while theirs does not. At $\theta
\neq 0$, their
 ${\bar{\cal F}}(\theta)$ is related to ${{\cal F}}(\theta)$ in (II.24) by
 \begin{equation}
 {\cal F}(\theta)=e^{i\pi{\Omega}}{\bar{\cal F}}(\theta) \ .
 \end{equation}
 Note that this matrix multiplication factor cannot be ignored when one
 considers
 the scattering cross section, ${d\sigma\over d\theta}|_{n_1n_2\rightarrow
n_3n_4}$,
 because of the effect of phase interference between the diagonalized channels.
  Also, in
 Ref.~\cite{lee}, they did not exchange  the particle labels $n_3$ and $n_4$ in
their
 exchange amplitude.

 The amplitude (II.24) depends on the momentum $p$ only through the kinematical
 factor, which reflects the conformal invariance of the system \cite{RJ2}. In
 fact, the action
 gotten by integrating eq. (II.1) possesses an $SO(2,1)$ conformal symmetry,
generated by
 time dilation
 \begin{eqnarray}
 t'&=&at \nonumber \\
 {\bf r}'&=&\sqrt{a} \br \nonumber \\
 \psi'(\br',t')&=&{1\over \sqrt{a}} \psi(\br,t) \nonumber \\
 A'_\mu(x')&=&{\partial x^\nu\over\partial {x'}^\mu} A_\nu (x) \ ,
 \end{eqnarray}
 conformal time transformation
 \begin{eqnarray}
 {1\over t'}&=&{1\over t}+a \nonumber \\
 {\bf r}'&=&{\br\over 1+ at} \nonumber \\
 \psi'(\br',t')&=&(1+at)e^{-{imr^2\over 2(1+at)}}\psi(\br,t) \nonumber \\
 A'_\mu(x')&=&{\partial x^\nu\over\partial {x'}^\mu} A_\nu (x) \ ,
 \end{eqnarray}
 and the usual time translation. This symmetry is broken however in
perturbation theory
 by quantum corrections, producing an anomaly.

\section{Perturbation Theory}
\renewcommand{\theequation}{\Roman{section}.\arabic{equation}}
\setcounter{equation}{0}

\hspace{0.5cm} We analyze the nonabelian AB scattering problem perturbatively
in a field
theoretic approach. We add to the Lagrange density (II.1) a gauge fixing term
 \begin{equation}
 {\cal L}_{gf}=-{1\over \xi}{\rm tr}\,(\nabla \cdot {\bf A})^2 \ ,
 \end{equation}
and a corresponding ghost term
\begin{equation}
 {\cal L}_{gh}=\eta^{*\,a}\left(\nabla^2 h_{ab} +
g f_{abc} {\bf A}^c \cdot\nabla\right)\eta^b \ .
 \end{equation}
The Feynman rules are derived from the total Lagrange density.
Fig.~(1) depicts the propagators of this theory, given in the limit
$\xi\rightarrow 0$
by
\begin{eqnarray}
D(p)&=&{i\over{p_0-{1\over{2m}}{\bf p}^2+i\epsilon}}  \\
G(p)&=&-{i\over{{\bf p}^2}} \\
G_{0i}(p)&=&-G_{i0}(p)={{\epsilon_{ij} p^j}\over{\kappa {\bf p}^2}} \\
G_{00}(p)&=&G_{ij}(p)=0\ .
\end{eqnarray}

\begin{figure}[htb]
\epsfxsize=5in
\centerline{\epsffile{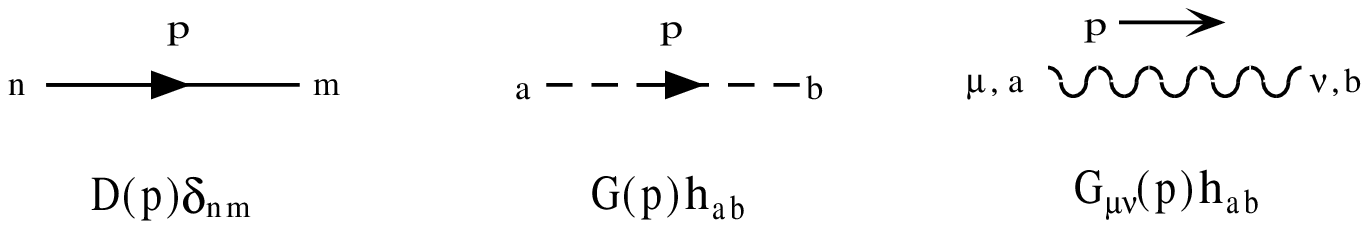}}
\vspace*{.1in}
\centerline{\small Figure~1.~ Propagators}
\end{figure}

\vspace*{.2in}

Fig.~(2) depicts the interaction vertices, given by
\begin{figure}[htb]
\epsfxsize=5in
\centerline{\epsffile{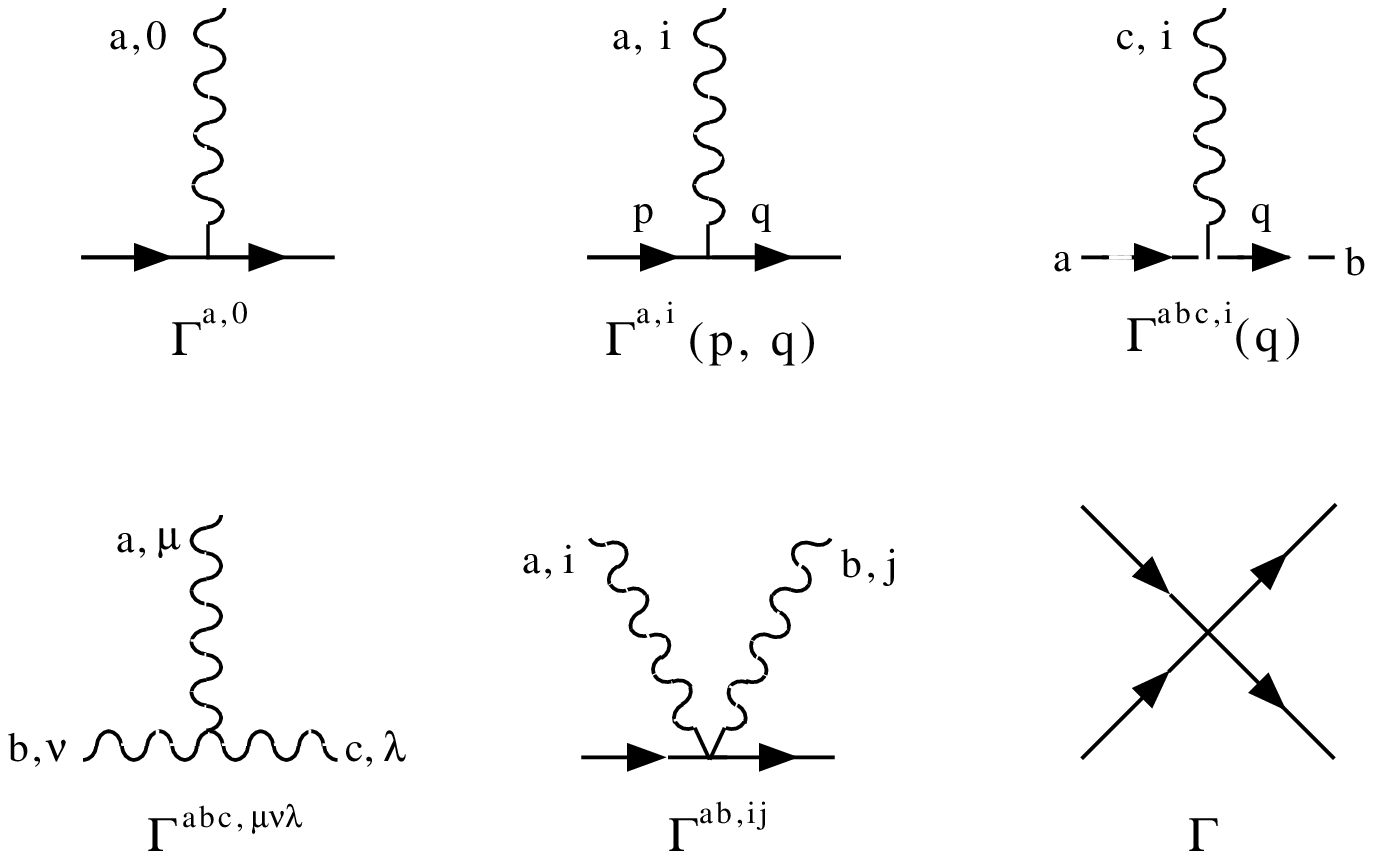}}
\centerline{\small Figure~2.~ Interaction Vertices}
\end{figure}

\goodbreak

\begin{eqnarray}
\Gamma^{a,0}&=&-gT^a \\ [.25cm]
\Gamma^{a,i}(p,q)&=&{g\over 2m}T^a(p^i+q^i) \\[.25cm]
\Gamma^{abc,i}(q)&=&-gf^{abc}q^i \\ [.25cm]
\Gamma^{abc,\mu\nu\lambda}&=&ig\kappa f^{abc}\epsilon^{\mu\nu\lambda} \\[.25cm]
\Gamma^{ab,ij}&=&{ig^2\over 2m} \bigg[T^aT^b + T^bT^a
\bigg]\delta^{ij}\\[.25cm]
\Gamma&=&{i\over 2}{\cal C} \ .
\end{eqnarray}

Before computing the scattering amplitude we need to check that there are no
corrections, at least to one loop, to the gluon propagator. These would
contribute unwanted divergences to the scattering amplitude. We already know
from the abelian theory that there are no corrections to the boson propagator,
 and we don't really care about the ghost propagator since it won't contribute
to the one loop boson 4-point function. Fig.~(3) depicts the two contributions
 to the gluon self energy, which only has space-space components.

\begin{figure}[htb]
\epsfxsize=3in
\centerline{\epsffile{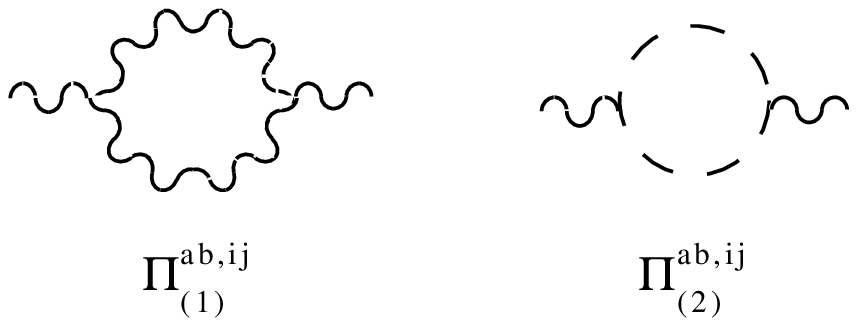}}
\centerline{\small Figure~3.~ Self Energy}
\end{figure}

\begin{eqnarray}
\Pi^{ab,ij}_{(1)}&=&{g^2\over 2}f^{acd}f^b_{\ cd}\dint{3}
 {k^i(k-p)^j-(i\leftrightarrow j)\over {\bf k}^2({\bf k}-{\bf p})^2}\\
\Pi^{ab,ij}_{(2)}&=& -g^2f^{acd}f^{b}_{\ cd}\dint{3} {k^i(k-p)^j\over {\bf
k}^2(
{\bf k}-{\bf p})^2} \ .
\end{eqnarray}
The total self energy is then
\begin{eqnarray}
\Pi^{ab,ij} &=& \Pi^{ab,ij}_{(1)}+\Pi^{ab,ij}_{(2)} \nonumber \\
&=& -{g^2\over 2}f^{acd}f^{b}_{\ cd}\dint{3} {k^i(k-p)^j-k^j(k-p)^i
  \over {\bf k}^2({\bf k}-{\bf p})^2} \nonumber \\
&=& 0 \ .
\end{eqnarray}

We compute the scattering amplitude by applying the Feynman rules to
calculate the 4-point function in the c.o.m. frame and multiplying the
result by $-i$. Fig.~(4) depicts the tree level contributions, resulting
in the amplitude :
\begin{equation}
{\cal A}^{(0)}={{\cal C}\over 2}-{i2\pi\over m}\Omega
  \cot{\theta\over 2} \ ,
\end{equation}
where $\theta $ is the scattering angle.

\begin{figure}[htb]
\epsfxsize=2in
\centerline{\epsffile{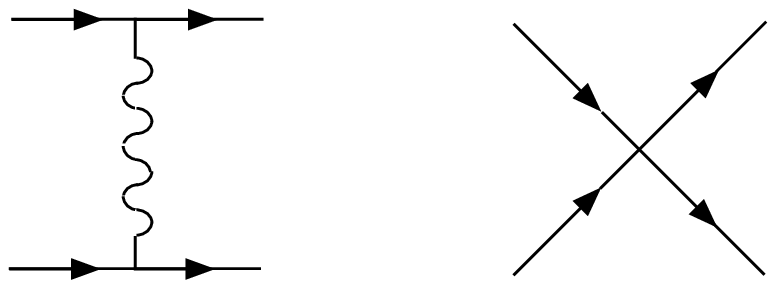}}
\centerline{\small Figure~4.~ Tree Level Scattering}
\end{figure}

Fig.~(5) depicts the one-loop contributions. Other than group matrix structure,
the new feature relative to the abelian theory is the tri-gluon diagram. At
first glance all the one loop contributions seem to be logarithmically
divergent. The
box diagram is finite however. Performing the $k_0$ integration yields
\begin{equation}
{\cal A}^{(1)}_{\rm box}({\bf p},{\bf p'})={16\pi^2\over {m}}\Omega^2 \dint{2}
{({\bf
k}\times{\bf p})({\bf k}\times{\bf p'})\over ({\bf k}+{\bf
p})^2({\bf k}+{\bf p'})^2(k^2-p^2-i\epsilon)}
\end{equation}
where ${\bf p}$ is the incident momentum in the c.o.m. frame, and ${\bf p}'$
 is the scattered momentum. Using the well known decomposition,
\begin{equation}
{1\over k^2-p^2-i\epsilon}={P\over k^2-p^2} +
i\pi\delta(k^2-p^2) \ ,
\end{equation}
we can split the amplitude into a real part and an imaginary part. The real
 part is given by

\begin{equation}
Re\left({\cal A}^{(1)}_{\rm box}(p,\theta)\right)=-{2\pi\over  m}\Omega^2
\ln|2\sin{\theta\over 2}| \ .
\end{equation}

\begin{figure}[htb]
\epsfxsize=3in
\centerline{\epsffile{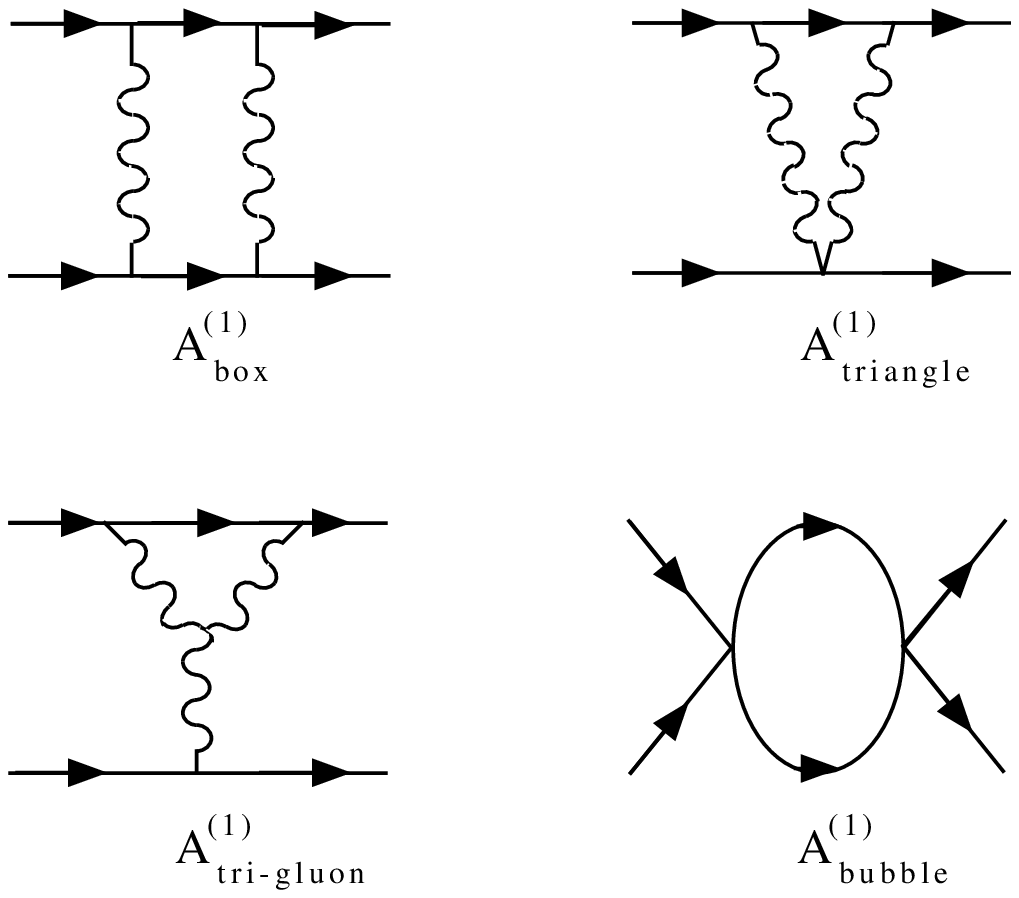}}
\centerline{\small Figure~5.~ One Loop Scattering}
\end{figure}

The computation of the imaginary part is somewhat subtle, but the result is
crucial. We expect a divergence in the forward direction on the grounds of
unitarity. Integrating over the angle and then taking the limit $k^2
\rightarrow p^2$ gives
\begin{equation}
Im\left({\cal A}^{(1)}_{\rm box}(p,\theta)\right)=-{2\pi^2\over m}\Omega^2
\left[1-2\pi\delta(\theta)\right] \ ,
\end{equation}
 reproducing the $\delta$-function of (II.24). In the field theoretic approach
one
implicitly assumes that the asymptotic states (incoming and outgoing) are
free particles, i.e. plane waves, so this result is consistent.

The triangle and tri-gluon contributions are given by
\begin{eqnarray}
{\cal A}^{(1)}_{triangle}&=&-{g^4\over 4m\kappa ^2}(T^aT^b\!+\!T^bT^a)\otimes
(T_aT_b)
\dint{2}{{\bf k}\cdot ({\bf k}-{\bf q})\over k^2({\bf k}-{\bf q})^2} \\
{\cal A}^{(1)}_{tri-gluon}&=&-{g^4\over 2m\kappa ^2}f_{abc}T^a\otimes (T^bT^c)
\dint{2}{k^2q^2-({\bf k}\cdot {\bf q})^2\over q^2k^2({\bf k}-{\bf q})^2} \ ,
\end{eqnarray}
where ${\bf q}\equiv {\bf p}-{\bf p}'$. Using
$$ T^aT^b+T^bT^a=2T^aT^b-[T^a,T^b]=2T^aT^b-f^{ab}_{\ \ c}T^c $$
we split $A^{(1)}_{triangle}$ in two parts, with different tensor structures,
\begin{eqnarray*}
{\cal A}^{(1)}_{triangle,1}&\propto & f_{abc}T^a\otimes \left( T_bT_c\right)\\
{\cal A}^{(1)}_{triangle,2}&\propto & \left( T^aT^b\right)\otimes \left( T_aT_b
\right) \ .
\end{eqnarray*}
By using Feynman reparameterization and Euclidean space dimensional
regularization we get
\begin{equation}
{\cal A}^{(1)}_{triangle,1}+{\cal A}^{(1)}_{tri-gluon}=0
\end{equation}
\begin{equation}
{\cal A}^{(1)}_{triangle,2}=-{\pi\over m}\Omega^2\left[
{1\over\epsilon}+\ln{4\pi\mu ^2\over p^2}-2\ln|2\sin{\theta\over 2}|-\gamma
+{\cal O}
(\epsilon)\right] \ ,
\end{equation}
where the dimension of space is taken to be $2-2\epsilon $, $\mu$ is an
arbitrary scale, and $\gamma$ is the Euler constant.
(At this point we note that without the contact term in the action the
theory would not be renormalizable, since there is no parameter to absorb the
$1/\epsilon$ divergence.)

The contribution of the bubble diagram is
\begin{eqnarray}
{\cal A}^{(1)}_{bubble}&=&{1\over 4}m{\cal C}^2\dint{2}{1\over
k^2-p^2-i\epsilon}
\nonumber \\
 &=&{1\over 16\pi}m{\cal C}^2\left[{1\over\epsilon}+\ln{4\pi\mu^2\over
p^2}+i\pi-\gamma
+{\cal O}(\epsilon )\right] \ ,
\end{eqnarray}
and the total one loop scattering amplitude is given by
\begin{equation}
{\cal A}^{(1)}={m\over 16\pi}\left[{\cal C}^2\!-\!{16\pi^2\over m^2}\Omega^2
\right]\left[{1\over\epsilon}\!+\!\ln{4\pi\mu ^2\over
p^2}\!+\!i\pi\!-\!\gamma\right]
+i{2\pi^3\over m}\Omega^2\delta (\theta)\ .
\end{equation}
This  amplitude is renormalized by redefining the contact interaction matrix
${\cal C}$ :
\begin{eqnarray}
 c(\alpha,\beta)&=&c_{ren}(\alpha,\beta)+\delta c(\alpha,\beta)\nonumber \\
 \delta c(\alpha,\beta)&=& -{m\over 8\pi}
\left({1\over\epsilon}\!+\!\ln{4\pi}\!-\!
\gamma\right)
\left[c^2_{ren}(\alpha,\beta)-{4g^4\over m^2\kappa
^2}\alpha^2\right]\nonumber\\
 {\cal C}_{ren}&=&\sum_{\alpha\beta w}
 |\alpha,\beta, w\rangle c_{ren}(\alpha,\beta)\langle\alpha,\beta, w| \ ,
\end{eqnarray}
and the total renormalized amplitude is given by
\begin{equation}
{\cal A}_{ren}(p,\theta,\mu)=-{2\pi i\over m}\left[{\Omega}
\cot ({\theta\over 2})\!+\!i{m{\cal C}_{ren}\over 4\pi}
\!-\!\Omega ^2\pi ^2 \delta(\theta)\right]+
{m\over 16\pi}\left({\cal C}_{ren}^2\!-\!{16\pi^2\over m^2}\Omega^2\right)
\left(\ln{\mu ^2\over p^2}\!+\!i\pi\right)\,.
\end{equation}
A conformal anomaly  appears through dependence
on an arbitrary scale.
There exist however critical points at which the
amplitude (III.28) is  conformally invariant, given by
\begin{eqnarray}
{\cal C}_{ren}^2-{16\pi^2\over m^2}\Omega^2 &=&0 \ .
\end{eqnarray}
Inserting (II.15) into (III.29), the solution in the diagonal basis is given
by
\begin{equation}
{\cal C}_{ren}=-{2g^2\over m|\kappa|}\sum_{\alpha\beta w}
\epsilon(\alpha,\beta)|\alpha, \beta, w\rangle|\alpha|
\langle\alpha, \beta, w| \ ,
\end{equation}
where $\epsilon(\alpha,\beta)$ is either $+1$ or $-1$ and does not depend on
$w$.
We still have the freedom to choose the sign in each
irreducible block of ${\cal C}_{ren}$. One solution
corresponds to choosing $\epsilon(\alpha, \beta)=\alpha/|\alpha|$ which gives
${\cal C}_{ren}=
-{2g^2\over m|\kappa|}T_a\otimes T^a={4\pi\kappa\over m|\kappa|}\Omega$.
Dunne et. al. have
found self-dual solitons for
this solution \cite{RJ1}. Another solution is gotten by choosing
$\epsilon(\alpha, \beta)=+1$,
resulting in a purely repulsive contact interaction in the diagonalized
two body Schr\"odinger equation.  For
the latter choice,
the total scattering amplitude is simply
\begin{equation}
{\cal A}(\theta)=-{2\pi i\over m}\left[{\Omega}
\cot ({\theta\over 2})-i|{\Omega}|
-\Omega ^2\pi ^2 \delta(\theta)\right] +{\cal O}({\Omega} ^3)\ ,
\end{equation}
which agrees, up to a kinematical factor, with the exact result in (II.24) to
 ${\cal O}(\Omega^3)$. Putting the matter indices back in
gives
\begin{equation}
A(n_1n_2\rightarrow n_3n_4,\theta)=\langle n_3n_4|{\cal
A}(\theta)|n_1n_2\rangle
+ \langle n_4n_3|{\cal A}(\theta +\pi)|n_1n_2\rangle\ .
\end{equation}
for the total scattering amplitude.

\section{Conclusion}
\renewcommand{\theequation}{\Roman{section}.\arabic{equation}}

The nonabelian AB scattering result is successfully obtained to one loop in
field theoretic perturbation theory. We demonstrated that contact interactions
are necessary for a renormalizable perturbation  theory, even though
they do not contribute in the exact treatment. The Schr\"odinger equation
(II.16)
requires physical input in the form of a boundary condition to obtain
an exact solution. Such a boundary condition cannot however be imposed in a
perturbative
treatment, but its physical content can be included in the form of a contact
interaction.

At critical values of the contact interaction, the theory is finite and
conformally
invariant. For a purely repulsive critical contact interaction, the
perturbative one loop
result agrees to second order with the exact solution with
vanishing boundary condition at the origin.

\begin{center}
{\bf ACKNOWLEDGMENTS}\end{center}
\ \indent
The authors would like to acknowledge Professor Roman Jackiw for
suggesting this problem. One of us (D. B.) also thanks
G. Amelino-Camelina for helpful discussions.

\end{document}